\documentclass[twocolumn,tighten]{aastex63}
\usepackage{amssymb, amsmath,framed}

\usepackage{bm}
\expandafter\ifx\csname package@font\endcsname\relax\else
 \expandafter\expandafter
 \expandafter\usepackage
 \expandafter\expandafter
 \expandafter{\csname package@font\endcsname}
\fi
\def\bq{\begin{equation}}
\def\eq{\end{equation}}
\def\bqy{\begin{eqnarray}}
\def\eqy{\end{eqnarray}}
\def\p{\partial}

\def\p{\partial}

\def\R{\mathbb{R}}

 \def\q#1#2{q^{\hspace{1pt}#1}_{\,,\hspace{1pt}#2}}
 \def\a#1#2{a^{\hspace{1pt}#1}_{\,,\hspace{1pt}#2}}
 \def\d#1#2{\delta^{\hspace{1pt}#1}_{\,,\hspace{1pt}#2}}

\begin{document}
\title{\large{Constraints on the abundance of $0.01\,c$ stellar engines in the Milky Way}}

\correspondingauthor{Manasvi Lingam}
\email{mlingam@fit.edu}

\author{Manasvi Lingam}
\affiliation{Department of Aerospace, Physics and Space Sciences, Florida Institute of Technology, Melbourne FL 32901, USA}
\affiliation{Institute for Theory and Computation, Harvard University, Cambridge MA 02138, USA}

\author{Abraham Loeb}
\affiliation{Institute for Theory and Computation, Harvard University, Cambridge MA 02138, USA}

\begin{abstract}
Stellar engines are hypothesized megastructures that extract energy from the host star, typically with the purpose of generating thrust and accelerating the stellar system. We explore the maximum potential speeds that could be realizable by stellar engines, and determine that speeds up to $\sim 0.1\,c$ might perhaps be attainable under optimal conditions. In contrast, natural astrophysical phenomena in the Milky Way are very unlikely to produce such speeds. Hence, astrometric surveys of hypervelocity stars may be utilized to conduct commensal searches for high-speed stellar engines in the Milky Way. It may be possible to derive bounds on their abundance, but this requires certain assumptions regarding the spatiotemporal distribution of such engines, which are not guaranteed to be valid. \\
\end{abstract}

\section{Introduction} \label{SecIntro}
The search for signatures of extraterrestrial technological intelligence (ETI) - appositely termed ``technosignatures'' by Jill Tarter in 2007 \citep{Tart07} - has been dominated by the quest for artificial electromagnetic signals from the 1960s onward, mostly at radio wavelengths \citep{Dra65,SS66,Tart01,WDS17,LBC20}; in spite of the numerous searches conducted hitherto, the fraction of parameter space sampled remains minuscule \citep{Tar10,WKL}. Since the very inception of this field, however, the potentiality of non-radio technosignatures was recognized, and its importance has been increasingly underscored in the 21st century \citep{Dys66,Fri83,BCD11,WMS14,Ci18,LL19}.

In classifying prospective ETIs, the Kardashev scale pioneered by the late Nikolai Kardashev has proven to be a valuable metric \citep{Kar64,Cir15,Gr20}. Type II ETIs, for instance, are capable of harnessing the entire radiative energy output of their host star. As these ETIs are considerably more advanced in terms of their technology than humans, it is conventionally anticipated that their resultant technosignatures would be commensurately more striking. The best known technosignatures in this category are Stapledon-Dyson spheres - these megastructures are composed of swarms of objects to tap the energy of the star \citep{Stap37,Dys60}. Several searches for Stapledon-Dyson spheres have been undertaken to date, as reviewed in \citet{Wri20} and \citet{LL21}.

Another group of megastructures belonging to a similar category are stellar engines, which draw upon the star's energy to extract useful work and typically generate thrust. Leonid Shkadov is widely credited with the first design for a stellar engine wherein a gigantic mirror was deployed to reflect a fraction of the radiation back toward the host star \citep{Shk87,Shk88}. However, \citet[pg. 260]{Zwi57} explicitly articulated this scenario in his characteristically wide-ranging monograph:
\begin{quote}
\emph{Considering the sun itself, many changes are imaginable. Most fascinating is perhaps the possibility of accelerating it to higher speeds, for instance $1000$ km/sec directed toward $\alpha$-Centauri in whose neighborhood our descendants then might arrive a thousand years hence. All of these projects could be realized through the action of nuclear fusion jets, using the matter constituting the sun and the planets as nuclear propellants.}
\end{quote}
Looking further back in time, the qualitative notion of stellar engines appears in \citet[Chapter XI]{Stap37}, as seen from the following quote:
\begin{quote}
\emph{The occasion of the first accident was an attempt to detach a star from its natural course and direct it upon an inter-galactic voyage \dots Plans were therefore made for projecting several stars with their attendant systems of worlds across the vast ocean of space that separated the two floating islets of civilization.}
\end{quote}
Stellar engines have been explored in several other publications \citep{BC00,BC06,DH18,Cap19,Svo20} and methods for detecting them during the course of exoplanetary transits were discussed in \citet{For13}.

In the classification scheme of \citet{BC00}, three major classes of stellar engines were identified. Class A stellar engines utilize the impulse from stellar radiation to generate a thrust force. The quintessential example of a Class A engine is the so-called Shkadov thruster described in \citet{Shk87,Shk88}. Class B stellar engines, in contrast, harness the radiation emitted by the host star and convert it into mechanical power. Class C stellar engines combine elements of both Class A and Class B stellar engines, and thereby generate both thrust force and mechanical power. Class D stellar engines, mentioned briefly in \citet[pg. 121]{BC06}, extract mass from the star by means of ``mass lifting'' \citep{Cri85} and expel the material to generate a rocket effect; such stellar engine were previously termed ``stellar rockets'' by \citet{Fogg}.

The designs suggested for the putative stellar engines vary from one class to another. The Shkadov thruster (Class A engine) is composed of a collection of reflective statites \citep{For91}, which effectively functions as a large mirror in the shape of a spherical arc; the design for a Class C engine is also similar in this respect \citep[Figure 1]{BC06}. Class D engines operate on the rocket effect, and can therefore be envisioned as stellar rockets, but the engineering specifics differ across proposals \citep{Cap19,Svo20}. We will briefly summarize one such design later in Sec. \ref{SSecSpeed}.

Clearly, stellar engines represent a massive engineering feat, and this raises the question of why they would be constructed. Before doing do, it should be noted that the planets and moons orbiting the host star would need to be accelerated as well, but this feat is relatively easier in comparison to the star. One possible reason suggested in \citet{Cap19} is that stellar engines may enable ETIs to preemptively escape the vicinity of catastrophic phenomena (e.g., supernovas) and avoid the adverse consequences.\footnote{For instance, theoretical models indicate that the habitability of the Milky Way might be affected over kpc scales due to the cumulative impact of tidal disruption events and a potential quasar phase at some point \citep{BT17,FL18,LoL19,LGB19,PBLT}.} This issue may be rendered more prominent near the Galactic center, where higher stellar densities are associated with higher rates of catastrophic events. Another option is that the ETIs might wish to undertake intergalactic travel, which is perhaps easier to undertake by moving the planetary system as a whole, in contrast to the alternative concept of building ``world ships'' \citep{HPP12}. We will not speculate on this topic further, as it partly overlaps with the poorly understood fields of xenosociology and xenopsychology.

In this work, we explore stellar engines from a generic physical standpoint in Sec. \ref{SecRat}, and describe how they may attain terminal speeds that are sub-relativistic. We compare these speeds against stars ejected by natural astrophysical phenomena in Sec. \ref{SecRat}, and argue that the latter cannot reach such large values in the Milky Way. By making use of this proposition, we examine current astrometric surveys to set tentative constraints on the abundance of putative ETIs that develop high-speed stellar engines in Sec. \ref{SecCSE}. We end with a summary of our results and prospects for future work in Sec. \ref{SecConc}.

\section{Maximum attainable stellar speeds}\label{SecRat}
We will begin with a synopsis of the maximal speeds realizable by ejected stars and stellar engines. 

\subsection{Maximum speeds of ejected stars in astrophysical systems}\label{SSecMaxS}
The maximum speed ($v_\mathrm{max}$) achievable by stars ejected after tidal disruption of a stellar binary system by Sagittarius A* was computed in the classic analysis by \citet{Hil88}, and it was concluded that $v_\mathrm{max} \approx 1.3 \times 10^{-2}\,c$; see also \citet{YT03} and \citet{Bro15}. Subsequent numerical simulations by \citet[Section 8]{SKR10} determined that the maximum speed of the least massive object in a triple system that underwent ejection is given by
\begin{equation}
    v_\mathrm{max} \approx 1.3 \sqrt{\frac{2 G M_2}{R_2 + R_3}} \left(\frac{M_1}{M_2 + M_3}\right)^{1/6},
\end{equation}
where $M_1$ denotes the mass of the supermassive black hole (SMBH), while $M_j$ and $R_j$ are the masses and radii of the stellar binary system that is subjected to tidal breakup ($j = 2,3$), where $M_3 < M_2$. We adopt $M_1 \approx 4 \times 10^6\,M_\odot$ for Sagittarius A* \citep{BGS16}, and substituting this value into the above equation yields
\begin{eqnarray}
   && v_\mathrm{max} \approx 3.4 \times 10^{-2}\,c\,\left(\frac{M_2}{M_\odot}\right)^{1/2}\left(\frac{R_2 + R_3}{R_\odot}\right)^{-1/2} \nonumber \\
   && \hspace{1.2in} \times \left(\frac{M_2 + M_3}{M_\odot}\right)^{-1/6},
\end{eqnarray}
and this expression is considerably simplified when $M_3$ is at least a few times smaller than $M_2$, and using the scaling $R_2 \propto M_2^{0.8}$ for main-sequence stars \citep{TPEH}. In this optimal situation, we end up with
\begin{equation}\label{vmaxMS}
v_\mathrm{max} \approx 3.4 \times 10^{-2}\,c\,\left(\frac{M_2}{M_\odot}\right)^{-0.07},
\end{equation}
which implies that $v_\mathrm{max}$ is nearly independent of $M_2$. Hitherto, we have assumed that $M_2$ is a main-sequence star, but the other extreme is to consider a stellar-mass black hole instead. The resultant maximum speed was calculated in \citet[Equation 5]{GL15}:
\begin{equation}\label{vmaxBH}
 v_\mathrm{max} \approx 6.7 \times 10^{-2}\,c\,\left(\frac{M_2}{10\,M_\odot}\right)^{1/6}\left(\frac{M_3}{M_\odot}\right)^{-0.12},
\end{equation}
where $M_2$ is the mass of a typical stellar-mass black hole and $M_3$ is the mass of the ejected star.

Thus, it is apparent from the preceding formulae that achieving $v_\mathrm{max} > 0.1\,c$ is very unlikely for gravitational triple interactions in our Galaxy. In fact, even ejected stellar speeds larger than $\sim 0.01\,c$ are rarely attained in numerical simulations, as seen from the probability distributions of ejected velocities in \citet{BKG,KBG08,GLW12,RKS14,GM20}. Faster speeds of $\gtrsim 0.1\,c$ are realizable in theory but require one of the stars in the stellar binary system to be replaced by a SMBH instead \citep{GL15,LG16,DCK19}; however, this specific scenario is manifestly not applicable to the Milky Way. 

Aside from the Hills mechanism and its variants mentioned earlier, it is necessary to gauge whether other avenues can eject stars at similar speeds. One of the most well-known processes entails the disruption of the binary when one of the objects undergoes a core-collapse supernova and leads to the ejection of the other objects \citep{Bla61,Boe61}. Numerical simulations indicate, however, that the maximal speeds of ejection are $\lesssim 0.01\,c$ \citep{Tau15,ERR20,Neu20}. Another possibility is the ``dynamical ejection scenario'' whereby dynamical ejection from stellar clusters is facilitated, typically at speeds of order $\lesssim 10^{-3}\,c$ \citep{PRA,Leo91,OK16}. Lastly, one could replace the SMBH with an intermediate mass black hole or a series of stellar-mass black holes, but the resulting speeds of ejected stars are $\lesssim 0.01\,c$ \citep{GZ07,OLL08,FG19}. In all these processes, the speeds attained are smaller than $v_\mathrm{max}$ for the Hills mechanism; see (\ref{vmaxMS}) and (\ref{vmaxBH}). 

\subsection{Stellar engines: potential speeds}\label{SSecSpeed}
The Shkadov thruster, which is an example of a Class A stellar engine, will reach the velocity $v$ after an interval $\Delta t$ as follows \citep[Equation 34]{BC00}:
\begin{equation}\label{Shk}
    v \approx 3.4 \times 10^{-5}\,c\,\left(\frac{\Delta t}{1\,\mathrm{Gyr}}\right)\left(\frac{L}{L_\odot}\right)\left(\frac{M_\star}{M_\odot}\right)^{-1},
\end{equation}
where $L_\star$ and $M_\star$ are the luminosity and mass of the star. In order to compare the speeds of stellar engines against those presented in Sec. \ref{SSecMaxS}, it is necessary to assess their \emph{maximal} values. 

It is reasonable to assume that the maximum speed is attained when $\Delta t \approx t_\star$, where $t_\star$ denotes the main-sequence lifetime of the star. From the scaling $t_\star \propto M_\star/L_\star$ \citep[Equation 1.90]{HKT04}, we find that $v_\mathrm{max}$ would become independent of stellar properties; after simplification, we obtain $v_\mathrm{max} \approx 3.4 \times 10^{-4}\,c$. Note, however, that this estimate for $v_\mathrm{max}$ applies only when $t_\star < t_U$, where $t_U$ is the current age of the Universe; this condition is fulfilled for $M_\star \gtrsim M_\odot$. On the other hand, if this criterion is violated, the stellar engine will not be able to achieve $v \approx v_\mathrm{max}$ in reality. Since the thrust generated by Class A and Class C engines is comparable \citep[Section 3]{BC06}, the same upper bound also applies to the latter.

Let us consider a generalized stellar engine wherein a fraction $\varepsilon$ of the total energy radiated by the star is harnessed to propel it, as suggested in \citet{DH18}. This would resemble a Class C stellar engine to a certain degree, because there is both energy extraction from the star's radiation and the generation of thrust force. By applying the conservation of energy, we obtain \citep[Equation 2.3]{DH18}:
\begin{equation}
    v \approx 1.2 \times 10^{-2}\,c\,\sqrt{\varepsilon}\left(\frac{\Delta t}{1\,\mathrm{Gyr}}\right)^{1/2}\left(\frac{L}{L_\odot}\right)^{1/2}\left(\frac{M_\star}{M_\odot}\right)^{-1/2},
\end{equation}
and we can calculate the maximum velocity by invoking the relation $\Delta t \approx t_\star$ from earlier, which yields
\begin{equation}\label{vmaxE}
    v_\mathrm{max} \approx 3.8 \times 10^{-2}\,c\,\sqrt{\varepsilon},
\end{equation}
and we reiterate that this speed is realizable in reality only when $M_\star \gtrsim M_\odot$. As opposed to energy conversion, if the momentum of radiation is harnessed, the scaling and magnitude of $v$ are akin to (\ref{Shk}).

We can, however, conceive of more sophisticated systems. Let us suppose, for instance, that instead of utilizing energy, the putative ETI extracts mass at a constant rate (denoted by $\dot{M}_\star$) via ``mass lifting'' \citep{Cri85}. This scheme does not exhibit a clear one-to-one mapping with any of the stellar engines described in Sec. \ref{SecIntro} because the focus is not on the energy or momentum of the emitted radiation, but rather on mass extraction and its subsequent conversion into thrust imparted to the star. The mass thus acquired is presumed to be converted into energy at an efficiency $\mu$ via the mass-energy relationship. In this event, provided that $\dot{M}_\star \Delta t \ll M_\star$ to ensure the mass is roughly constant, the speed achieved over the interval $\Delta t$ is estimated to be
\begin{equation}
    v = \sqrt{\frac{2 \mu c^2 \dot{M}_\star \Delta t}{M_\star}}.
\end{equation}
In place of working with two variables in the above expression, we can rewrite it as follows: we suppose that ETI modulates its mass extraction and energy conversion such that $\dot{M}_\star \Delta t = \zeta M_\star$ with $\zeta \ll 1$, i.e., the ETI ensures that the star's mass is not significantly depleted after the interval $\Delta t$.\footnote{The depletion of stellar mass, among other things, may aid in controlling the stellar luminosity, and thereby mitigating the shift of the habitable zone around the star over time.} A crude upper bound on $\zeta$ is $0.1$ (larger values would result in non-negligible stellar mass loss), which consequently yields a maximum speed of
\begin{equation}\label{vmaxME}
    v_\mathrm{max} \approx 4.5 \times 10^{-1}\,c\,\sqrt{\mu}\,\left(\frac{\zeta}{0.1}\right)^{1/2}.
\end{equation}
It is important to note, however, that both energy and momentum of the system as a whole are conserved. In the latter case, we would have $M_\star v \approx (\zeta \mu  M_\star c^2)/c$ by momentum conservation, thus obtaining
\begin{equation}\label{vmaxMP}
    v_\mathrm{max} \approx 0.1\,c\,\mu\,\left(\frac{\zeta}{0.1}\right),
\end{equation}
in the same manner as (\ref{vmaxME}); for $\mu \ll 1$, it is apparent that this velocity drops below that of (\ref{vmaxME}). Due to the joint conservation of energy and momentum, it follows that the peak velocity cannot exceed the maximum of (\ref{vmaxME}) and (\ref{vmaxMP}) for this generic stellar engine.

It is necessary to examine the potential values of $\mu$ further. In the canonical case of the proton-proton chain reaction to yield helium, it is well-known that $\mu \approx 0.7\%$. Thus, it would appear as though (\ref{vmaxME}) is comparable to (\ref{vmaxE}) \emph{prima facie}. It is worth mentioning, however, that efficiencies of $< 42\%$ are predicted for black holes by general relativity \citep{NT73}. Thus, if the stellar engine is a binary with one of the objects being a black hole, it is conceivable that $\mu \sim 10\%$ could be effectuated, although the subsequent pathway to imparting thrust to the binary system remains indeterminate (and undoubtedly complicated) and falls outside the scope of this paper. What is clear, however, is that the mass extraction can continue as long as access to the black hole is maintained, which implies that the black hole must also be accelerated. Even setting aside this option, we remark that ETIs with the technological wherewithal to build stellar engines might find methods to raise $\mu$ by an order of magnitude or so compared to the proton-proton chain. In case $\mu \sim 0.1$ can be effectuated through a suitable avenue, the above formula leads to $v_\mathrm{max} \sim 0.1\,c$. 

The above analysis, however, ignored the fact that a continual mass loss ought to result in a rocket effect. In theory, we can also conceive of thrusters propelled by jets in stellar and compact object systems, corresponding to the Class D stellar engines mentioned in Sec. \ref{SecIntro}. The final velocity is straightforward to calculate when the rocket equation holds true \citep{KT03}:
\begin{equation}
    v = v_\mathrm{ex} \ln\left(\frac{M_i}{M_f}\right),
\end{equation}
where $M_i$ and $M_f$ are the initial and final masses of the star, whereas $v_\mathrm{ex}$ is the velocity at which the propellant is expelled. If one considers $v_\mathrm{ex}$ that is a few times higher than the stellar escape velocity, or equivalently the stellar wind velocity, and choose a mass ratio of $\sim 10$,\footnote{Note, however, that this mass ratio (which is equivalent to $\zeta \sim 0.9$) is at odds with the choice of $\zeta$ considered previously. We have considered a deliberately high value to highlight the significance of the rocket effect when it comes to Class D stellar engines.} we would end up with $v \sim 0.01\,c$. It should be noted that this setup calls for an increase in the exhaust velocities by only a factor of order unity compared to current designs \citep{CCS15,WW16}. From a conceptual standpoint - although they are not readily implementable with current human technology - relativistic rockets reliant on hydrogen (or its isotopes) as the fuel have the capacity to reach  weakly relativistic exhaust velocities \citep{Win19,HZG20}; in this context, the realization of $v_\mathrm{max} \sim 0.1\,c$ does not seem wholly impossible.

We point out that two recent designs for Class D stellar engines have examined the aforementioned phenomena in more detail \citep{Cap19,Svo20}. In the so-called ``Star Tug'' proposed in \citet[Figure 1]{Svo20}, mass lifting is used to extract matter from a Sun-like star, which is converted into propellant by an engine located at a given distance. This propellant is employed to generate thrust that overcomes the gravitational force between the engine and the star, and causes the acceleration of the system. At perfect efficiency, the Star Tug achieved $v_\mathrm{max} \approx 0.27\,c$ in a few Gyr, whereas lowering the efficiency to $20\%$ still enabled a velocity of $0.1\,c$ to be achieved in approximately $10$ Gyr. The Star Tug achieved an asymptotic acceleration of $\sim 10^{-7}$ m s$^{-2}$ at perfect efficiency when the engine was situated far away from the star, but this value dropped by nearly two orders of magnitude at $20\%$ efficiency.

Once the stellar engine has begun accelerating and its passage through the interstellar medium (ISM) is underway, it may be feasible to make use of other propulsion systems to generate additional thrust and couple them to the stellar engine. The interstellar ramjet, which scoops up interstellar material and converts the accrued matter into fuel, represents one such possibility \citep{Bus60,Long11}. We will not delve into the technical details of this putative coupling scheme, because ascertaining the engineering technologies adopted by hypothetical advanced ETIs is indubitably constrained by our current level of knowledge and vision.

Lastly, a comment on the effects of the ISM on stellar engines is in order. The mechanical components comprising the stellar engine will be subject to erosion and damage due to interactions with gas, dust and cosmic rays in the ISM. However, predicting the extent of damage is difficult because it depends on the relative velocity of the stellar engine, the distance travelled by it, the thickness and properties of the materials used to fabricate the components, and many more. There have been several theoretical studies of the passage of sub-relativistic spacecraft through the ISM, and it might be feasible for them to remain functional over kpc distances \citep{HLBL,HL17,Ho17,HLL,LiMa20}, although significant uncertainties still remain.

To summarize, there appear to be plausible designs for stellar engines that seem capable of achieving $v_\mathrm{max} \sim 0.01$-$0.1\,c$ in principle. At the minimum, it may be argued that no compelling \emph{a priori} grounds exist for dismissing the prospects for sub-relativistic stellar engines based on core physical principles.

\section{Constraints on stellar engines}\label{SecCSE}
In the preceding Section, we presented arguments as to why even the fastest stars ejected from astrophysical systems are unlikely to have speeds of $\gtrsim 10^4$ km/s ($0.03\,c$), whereas stellar engines could attain such velocities. Hence, if one were to detect stars moving at $\gtrsim 0.1\,c$, it would be strongly indicative of ETI activity and thus constitute a technosignature. The one false positive that ought to be taken into consideration is that a star moving at $\gtrsim 0.1\,c$ may have an extragalactic origin because of ejection during binary SMBH interactions \citep{GL15}, as explained in Sec. \ref{SSecMaxS}. A combination of precise astrometry and chemical tagging should, however, aid in distinguishing between extragalactic and Galactic hypervelocity stars. This procedure was utilized to pinpoint the origin of HVS 3 from the Large Magellanic Cloud \citep{EBG19}.

The \emph{Gaia} mission was designed to accurately pin down the positions and radial velocities of $\sim 10^9$ stars in the Milky Way. \emph{Gaia} DR-1 and DR-2 have already yielded a wealth of data on this front \citep{Gaia1,Gaia2}. As a rigorous assessment of the origin of hypervelocity stars requires knowledge of their total velocity \citep{Bro15}, it is necessary to measure both their radial and tangential components. \emph{Gaia} DR-2 has provided radial velocity information about $7$ million stars \citep{KSC19}. The number of stars with data available regarding their proper motion (i.e., tangential velocity) is $\sim 1.3 \times 10^9$ \citep{LHB18}. A number of studies have already combed through this sample to unearth evidence for hypervelocity stars \citep{BKB18,HVB18,MRB19,BSA19,CMB20,LJG20}. 

To the best of our knowledge, the fastest hypervelocity star from \emph{any} survey is S5-HVS1 from the Southern Stellar Stream Spectroscopic Survey with a velocity of $\sim 6 \times 10^{-3}\,c$ \citep{KBL20}, while HVS 22 from the Multiple Mirror Telescope survey exhibits a similar velocity of $\sim 5 \times 10^{-3}\,c$ \citep{KIH20}. Among the \emph{Gaia} DR-2 catalog, one of the fastest objects is the candidate \emph{Gaia} DR-2 6097052289696317952, which appears to have a tangential velocity of $\sim 5.4 \times 10^{-3}\,c$ \citep{Sch18}. Other analyses of the \emph{Gaia} DR-2 data have seemingly identified stars with velocities (tangential and/or radial) that are comparable to, but somewhat lower, than this object \citep{SBG18,BKB18,DLY19,CMB20}.

Thus, it would seem a safe bet to contend that all searches conducted to date have failed to yield unambiguous evidence of sub-relativistic stellar engines. This datum enables us to derive the following constraint:
\begin{equation}\label{DrakeSE}
    N_\mathrm{surv} \cdot f_\mathrm{T} \cdot f_\mathrm{SE} < 1,
\end{equation}
where $N_\mathrm{surv}$ represents the number of stars collectively encompassed by all astrometric surveys, $f_\mathrm{T} $ denotes the fraction of all stars that host ETIs with human-level technology, and $f_\mathrm{SE}$ embodies the fraction of all human-level ETIs that subsequently achieve the \emph{capability and intent} to deploy sub-relativistic stellar engines. The reason we select human-level technology as the benchmark is not due to the notion that we are ``special'', but because it constitutes a signpost that is familiar to us.

Although (\ref{DrakeSE}) may appear simple, the estimation of the sample size $N_\mathrm{surv}$ is fraught with several difficulties and ambiguities, some of which are detailed below.
\begin{enumerate}
    \item Given that \emph{Gaia} DR-2 furnished astrometric data for $\sim 10^9$ stars and radial velocity data for $\sim 10^7$ stars, it is tempting to specify $N_\mathrm{surv} \sim 10^7-10^9$.
    \item However, the stellar engines we discussed in Sec. \ref{SecCSE} reach their peak speeds of $\sim 0.01$-$0.1\,c$ only when the stars have masses close to that of the Sun or higher ($M_\star \gtrsim M_\odot$). Among the total population of $\sim 10^9$ objects spanned by the \emph{Gaia} mission, only $\sim 10^8$ of them are solar-type main-sequence stars.\footnote{\url{https://www.cosmos.esa.int/web/gaia/faqs}} Hence, the above numbers for $N_\mathrm{surv}$ must be reduced by roughly an order of magnitude when we restrict ourselves to solar-type stars.
    \item  The calcium triplet (CaT) is used by the \emph{Gaia} mission to estimate the radial velocities by measuring the Doppler shift. The spectrograph has a bandpass of $847$-$874$ nm \citep{CKS18}, thereby providing a ``buffer'' of $\sim 8$ nm because the longest CaT spectral line is manifested at $866.2$ nm. Stellar engines moving at radial velocities of $\sim 0.01$-$0.1\,c$ would yield Doppler shifts of $\sim 8.5$-$85$ nm at the CaT wavelength(s). Thus, the \emph{Gaia} spectrograph is liable to ``reject'' such stellar engines when the shift falls outside its bandpass. In principle, however, high tangential velocities can be inferred from measurements of the proper motions and parallaxes \citep{SBG18,Sch18}.
    \item Last, but not least, (\ref{DrakeSE}) is expected to yield an accurate picture only when the stellar engines are uniformly distributed in both the spatial and temporal realms. However, it will take $\lesssim 1$ Myr for the stellar engines to exit the $\sim 1$ kpc region spanned by \emph{Gaia} DR2 once their maximal velocity is attained, and they will eventually cross the entire Milky Way in $\sim 10$ Myr. Hence, it is not straightforward \emph{a priori} to derive proper constraints on the unknown parameter(s) unless new stellar engines are being continually ``injected'' into the domain encompassed by surveys like \emph{Gaia}.
\end{enumerate}

Bearing these caveats in mind, let us adopt a rough estimate of $N_\mathrm{surv} \sim 10^6$ with the express purpose of taking the argument further; we emphasize that this value is fiducial. By plugging this choice into (\ref{DrakeSE}), we obtain
\begin{equation}
    f_\mathrm{T} \cdot f_\mathrm{SE} < 10^{-6} \left(\frac{N_\mathrm{surv}}{10^6}\right)^{-1}.
\end{equation}
After the \emph{Gaia} mission is complete, assuming that no stellar engines are found, we may end up with the potentially tightest limit of $f_\mathrm{T} \cdot f_\mathrm{SE} < 10^{-8}$ for speeds $\gtrsim 0.01\,c$ under optimal circumstances, i.e., when the subtleties described in points \#3 and \#4 are set aside. As these constraints from surveys might prove to be quite stringent, the implications are elucidated below.
\begin{itemize}
    \item The first possibility is that $f_\mathrm{T}$ is exceptionally small. In this case, the prevalence of human-level ETIs would be commensurately low. This could arise due to any number of evolutionary bottlenecks ranging from abiogenesis to complex multicellularity to technological intelligence \citep{MSS95,LL21}.
    \item In the second case, $f_\mathrm{SE}$ may be minuscule, but $f_\mathrm{T}$ may have a moderate magnitude. There are, however, different scenarios at play here. On the one hand, ETIs might have a short technological lifetime and become extinct before they reach the stage where they can build stellar engines. On the other hand, it could very well be that weakly relativistic stellar engines have hidden engineering obstacles that render their construction impossible, or that ETIs possess the capability to build them but opt not to do so for other reasons.
    \item In the third outcome, both $f_\mathrm{T}$ and $f_\mathrm{SE}$ are both very small. This situation does not warrant separate explication, because it represents an amalgamation of the above two points.
\end{itemize}
To differentiate between, and indeed shed light on, the diverse outcomes demarcated above, the practical importance and necessity of carrying out searches for biosignatures and technosignatures on multiple fronts is self-evident \citep{Fra18,HKS20}.

There is a fourth option that deserves to be mentioned at this juncture. In theory, it is possible that stellar engines might already exist in the Milky Way, but that these putative megastructures are operating at velocities that fall \emph{within} the bounds of known hypervelocity stars. In this setting, distinguishing between them and naturally occurring hypervelocity stars would be an extremely challenging endeavour. We will not explore this scenario further because it calls for additional assumptions about the preferred trajectories of stellar engines, and this requires an understanding of the motives of putative ETIs, which is wholly unknown.

Before moving ahead, we note that stellar engines are potentially capable of accelerating at $\sim 10^{-9}$ m s$^{-2}$ \citep{Cap19,Svo20}. In contrast, the centripetal acceleration of the Sun is $\sim 10^{-10}$ m s$^{-2}$. Thus, in principle, detecting anomalously high stellar accelerations might also be indicative of stellar engine activity, although we caution that the ratio of the two accelerations (i.e., an order of magnitude) is not strikingly large.

\section{Conclusion}\label{SecConc}
We examined various mechanisms for the ejection of stars at high speeds, and concluded that stars ejected in the Milky Way are very unlikely to attain speeds over $\gtrsim 10^4$ km/s ($0.03\,c$) by any known natural astrophysical phenomena. Next, we considered some proposed designs for stellar engines, i.e., propulsion systems engineered by advanced ETIs to accelerate stars, which were conceived by Olaf Stapledon and Fritz Zwicky (among others) in the mid-20th century. We argued that speeds of $\sim 0.01$-$0.1\,c$ may be potentially achievable by stellar engines under optimal circumstances.

Based on the above premises, we examined current surveys for hypervelocity stars including the recent \emph{Gaia} DR-2 sample. In light of existing studies, we concluded that no stars have been conclusively identified that possess velocities of $\gtrsim 0.01\,c$. Taken at face value, stellar engines moving at sub-relativistic speeds appear to be quite rare but placing stringent constraints on their likelihood is rendered difficult because of both instrumental limitations and lack of knowledge about their spatio-temporal distribution. It might be possible that fewer than one in $\sim 10^6$ stars is propelled to speeds $\gtrsim 0.01\,c$ by stellar engines, although this statement is not definitive in light of the attendant uncertainties.

In the future, it would seem worthwhile to pursue the search for stars with anomalously high velocities (namely $\gtrsim 0.03\,c$). This strategy is advantageous for two chief reasons. First, it does not necessitate any new resources, because it can readily piggyback on astrometric surveys like \emph{Gaia}. Second, if we do stumble upon stars moving at such anomalously high speeds, their origin would be of great interest and significance irrespective of whether they have an artificial basis or not. 

In this regard, this search exemplifies the philosophy underpinning the aptly named `` First Law of SETI Investigations'' proposed by the late Freeman Dyson: ``\emph{Every search for alien civilizations should be planned to give interesting results even when no aliens are discovered.}'' \citep{Tech18}, which itself echoes the earlier sentiments espoused by Frank Drake in his neglected early treatise \citep[pg. 342]{Dra65}: ``\emph{Thus, any project aimed at the detection of intelligent extraterrestrial life should simultaneously conduct more conventional research.}''\footnote{On a broader note, the significance of seeking ``anomalies'' \emph{sensu lato} is garnering deeper appreciation in the adjacent realm of biosignatures as well \citep[Chapter 8]{Cle19}.} Thus, in each of these respects alongside a few others, this search for technosignatures may score highly on the axes of merit adumbrated in \citet{Sof20}.

\acknowledgments
We are grateful to the two reviewers for their positive and insightful reports, which helped improve the paper. This work was supported in part by the Breakthrough Prize Foundation, Harvard University's Faculty of Arts and Sciences, and the Institute for Theory and Computation (ITC) at Harvard University.

%\bibliographystyle{aasjournal}
%\bibliography{Thruster}

\end{document}